\documentclass[]{svjour3}

\usepackage{amsmath}
\usepackage{amssymb}
\usepackage{mathptmx}
\usepackage{isomath}

\usepackage{comment}
\usepackage{graphicx}

\newcommand{\bm}[1]{\vec{#1}}

\newcommand{\ci}{\mathrm{i}}
\newcommand{\ce}{\mathrm{e}}
\newcommand{\cpi}{\pi}
\newcommand{\dd}{\mathrm{d}}
\newcommand{\abs}[1]{\lvert #1 \rvert}
\newcommand{\average}[1]{\langle #1\rangle}
\newcommand{\ck}{k_{\text{B}}}
\newcommand{\Tr}{\operatorname{Tr}}
\newcommand{\cconj}[1]{{#1}^{*}}

\begin{document}

\title{Emergence of Non-Axisymmetric Vortex in Strong-Coupling Chiral \textit{p}-Wave Superconductor}
\author{Noriyuki Kurosawa \and Yusuke Kato}
\institute{N. Kurosawa \and Y. Kato \at Department of Basic Science, The University of Tokyo, Komaba 3-8-1 Meguro Tokyo 153-8902, Japan, \\
\email{kurosawa@vortex.c.u-tokyo.ac.jp}\\\email{yusuke@phys.c.u-tokyo.ac.jp}}

\maketitle

\begin{abstract}
We studied strong-coupling effect upon an isolated vortex in a two-dimensional chiral \textit{p}-wave superconductor. We solved the Eilenberger equation for the quasiclassical Green's functions and the \'{E}liashberg equation with single mode Einstein boson self-consistently. We calculated the free-energy of each obtained vortex, and found that a non-axisymmetric vortex metastably exists in some situation.
\end{abstract}

\section{Introduction}
The Migdal and \'{E}liashberg theory of ``strong-coupling superconductivity''\cite{Migdal1958,Eliashberg1960,Morel1962} has been very successful in qualitative description of superconductivity in real materials\cite{Scalapino1969,McMillan1969,Carbotte1990}.
For example, it explains the deviation from the universal value \((2\Delta)/(\ck T_{\text{c}})\approx3.53\) in Bardeen-Cooper-Schrieffer (BCS) theory, or the dependence of critical magnetic field on the transition temperature.
Moreover, in some situation, it is known that the strong-coupling effect is not only a quantitative but also a qualitative effect (e.g., Refs.~\cite{Marsiglio1991,Combescot1995}).
The strong-coupling effect modifies the spectrum of the quasiparticles. Therefore, one can expect that it may change the structure of low-energy states within the vortices in type-II superconductors. As far as we know, however, there have been only a few studies of the strong-coupling effect on a vortex on the basis of microscopic theories.

Two-dimensional chiral \textit{p}-wave superconductivity is considered to be realized in Sr\(_2\)RuO\(_4\) \cite{Maeno1994,Mackenzie2003,Sigrist2005,Maeno2012}. This state is topologically non-trivial and attracts much attention in these days. Within the vortices of this superconductor, reflected in the topology of this system, there is a zero-energy bound state, which is expected to be very robust against not-so-strong impurities\cite{Volovik1999,Matsumoto2000,Kato2000,Hayashi2005,Tanuma2009,Eschrig2009,Kurosawa2013,Kurosawa2015,Tanaka2016}.
Recently, the relationship between this robustness and the odd-frequency pairing also has been discussed\cite{Tanuma2009,Tanaka2016,Tanaka2012}.

In the present paper, we calculated the self-consistent \'{E}liashberg equation to study how the strong-coupling feature affects the vortex of a chiral \textit{p}-wave superconductor microscopically. We also calculated the free-energies of the vortices and discuss its stability.

\section{Methods}
In this study, we consider an isolated vortex in the two-dimensional spinless chiral \textit{p}-wave superconductor with isotropic Fermi surface.
%We assume that the London pentration depth is much longer than the coherent length of the superconductor (type-II limit) and ignore the electromagnetic potentials for simplisity.
We use quasiclassical theory\cite{Eilenberger1968,Larkin1968}; we assume that the product of the coherence length of the superconductor \(\xi\) and the Fermi wavevector \(\bm{k}_{\text{F}}\) is much larger than unity. The quasiclassical Green's function \(\check g\) is a \(2\times 2\) matrix and obeys the Eilenberger equation
\begin{align}
  \ci\hbar\bm{v}_{\text{F}}\cdot\nabla \check g(\ci\epsilon_n,\alpha,\bm{r}) + \left[\ci\hbar\epsilon_n\check\tau_3 + (q/c)\bm{v}_{\text{F}}\cdot\bm{A}\check\tau_3 - \check\Sigma(\ci\epsilon_n,\alpha,\bm{r}) ,\; \check g(\ci\epsilon_n,\alpha,\bm{r})\right]&= \check 0
  \label{Eq:Eilenberger}
  ,
\end{align}
where \(\epsilon_n = (2n+1)\cpi \ck T/\hbar\) are the Matsubara frequencies, \(\bm{v}_{\text{F}}\) is a Fermi velocity, \(\alpha\) denotes a direction of momentum on the Fermi surface such that \(\bm{k} = k_{\text{F}}(\cos\alpha, \sin\alpha)\), \(\check\tau_i\) (\(i=0,\dots 3\)) are the Pauli matrices, \(q\) is the elementary charge, \(c\) is the speed of light, \(\bm{A}\) is the vector potential, and
\begin{align}
\check\Sigma &= \begin{pmatrix}
\sigma & \Delta \\ -\cconj{\Delta} & -\sigma
\end{pmatrix}
,
\end{align}
is the self-energy.
The quasiclassical Green's function satisfies the normalization condition
\(\check g^2 = -\cpi^2\check\tau_0\)
and its bulk value is
\begin{align}
\check g = \frac{\cpi}{\sqrt{-(\ci\hbar\epsilon_n-\sigma)^2+\abs{\Delta}^2}}
\begin{pmatrix}
-\ci\hbar\epsilon_n + \sigma & \Delta \\
-\cconj{\Delta} & \ci\hbar\epsilon_n - \sigma
\end{pmatrix}
.
\end{align}

To incorporate strong-coupling effect, we use \'{E}liashberg equation to calculate the self-energy \(\Sigma\) from the quasiclassical Green's function;
\begin{align}
  \check\Sigma(\ci\epsilon_n, \alpha, \bm{r})
  &=
  N_0\ck T\sum_{\epsilon_m}^{\abs{\epsilon_m} < \epsilon_{\text{c}}}\average{v(\ci\epsilon_n, \alpha, \ci\epsilon_m, \alpha')\check g(\ci\epsilon_m, \alpha', \bm{r})}_{\alpha'}
  \label{Eq:Eliashberg}
  ,
\end{align}
where \(\epsilon_{\text{c}}\) is the cutoff of the Matsubara frequencies, \(N_0\) is the density of states on the Fermi level, and \(\average{\dots}_{\alpha}\) denotes the average over the Fermi surface and is defined \( \average{A(\alpha)}_{\alpha} = \int_0^{2\cpi}\dd\alpha A(\alpha)/(2\cpi)\). We took 47 equally spaced points in the momentum space. We assume that the interaction between electrons \(v\) has the following form:
\begin{align}
  v(\ci\epsilon_n,\alpha,\ci\epsilon_m,\alpha')
  &=
  \frac{C\omega_0^2}{(\epsilon_n-\epsilon_m)^2 + \omega_0^2} \times 2\cos(\alpha-\alpha')
  ,
\end{align}
where \(\omega_0\) is a characteristic frequency of a mediated boson, and \(C\) is a constant parameter. We set these parameters so that \(\hbar\omega_0 = 3\ck T_{\text{c}}\), where \(T_{\text{c}}\) is the critical temperature of the superconductivity. We choose this value so that the strong-coupling effect is very large but not unrealistic\footnote{With this parameter, the ratio of the energy gap to the critical temperature (\(2\Delta/T_{\text{c}}\)) is about \(5.6\) at \(T=0.02T_{\text{c}}\). For example, CeCoIn$_5$ exhibits such a large value (\(\approx 6\)) \cite{Park2008}.}. We set the cutoff of the Matsubara frequencies \(\hbar\epsilon_{\text{c}} = 20\ck T_{\text{c}}\), and confirm that this cutoff is considered to be sufficiently large by comparison of the magnitude of the bulk pair-potential with those for \(\hbar\epsilon_{\text{c}} = 10\ck T_{\text{c}}\) and \(15\ck T_{\text{c}}\). We also define \(\xi_0 = \hbar v_{\text{F}}/(\ck T_{\text{c}})\) and use it as a characteristic length of the spatial modulation of the self-energy.

The vector potential \(\bm{A}\) is obtained from the quasiclassical Green's function as
\begin{align}
  \nabla\times(\nabla\times\bm{A}) &= \frac{q}{c}N_0\ck T\sum_{\epsilon_n}\operatorname{Tr}\average{\bm{v}_{\text{F}}\check g_{11}}
  ,
\end{align}
where \(\Tr\) is a trace over the Nambu space.
We define \(\lambda_0 = (N_0v_{\text{F}}^2q^2c^{-2})^{-1/2}\) as a characteristic length of the electromagnetic entities. We set \(\lambda_0/\xi_0 = 2.5\) in this paper.

To discuss the stability of the isolated vortices, we calculated the free-energy deviation from the normal state \(\Omega_{\text{sn}}\) with the following equation:
\begin{align}
  \Omega_{\text{sn}}
  &=
  \int\dd\bm{r}\left(N_0 \ck T\sum_{\epsilon_n}\Tr\left\{\int_0^1\dd s\average{\check g_s\check\Sigma}-\frac{1}{2}\average{\check g\check\Sigma}\right\} + \frac{B^2}{2}\right)
  \label{Eq:free-energy-super-normal}
  ,
\end{align}
 where \(\bm{B} = \nabla\times\bm{A}\) is the magnetic field, and \(\check g_s\) is a solution of
\begin{align}
  \ci\hbar\bm{v}_{\text{F}}\cdot\nabla \check g_s + \left[\ci\hbar\epsilon_n\check\tau_3 + (q/c)\bm{v}_{\text{F}}\cdot\bm{A}\check\tau_3 - s\check\Sigma ,\; \check g_s\right]&= \check 0
  , &\check g_s^2=-\cpi^2\check\tau_0.
\end{align}
The above expression of \(\Omega_{\text{sn}}\) is a simple extension of the weak-coupling BCS one\cite{Serene1983,Thuneberg1984}. We used the 15-points Gauss-Kronrod quadrature formula to integrate respect to \(s\).
%We neglect the boson part of free-energy\cite{Lee1988}.

We numerically confirmed that the self-energy for Matsubara-frequencies \(\check\Sigma(\ci\epsilon_n, \alpha, \bm{r})\) can be decoupled as
\begin{align}
  \check\Sigma(\ci\epsilon_n,\alpha,\bm{r}) &=
  h(\ci\epsilon_n)\left[\check\Sigma_{+}(\bm{r})\ce^{+\ci\alpha} +\check\Sigma_{-}(\bm{r})\ce^{-\ci\alpha}\right]
  \label{Eq:self-energy-form}
  ,
\end{align}
and thus we show only the \(\bm{r}\)-dependent part \(\check\Sigma_{+}(\bm{r})\) and \(\check\Sigma_{-}(\bm{r})\) in the following section.
At sufficiently far from the vortex, only \(\check\Sigma_{+}\) or \(\check\Sigma_{-}\) survives. Hereafter we assume that \(\check\Sigma_{+}\) is a dominant part of the self-energy and survives in the bulk.
As we can see in \eqref{Eq:self-energy-form}, Cooper pair of chiral \textit{p}-wave superconductivity has internal angular momentum (chirality). If there is a vortex, two types of vortices can exist in this system; one type of vortex has vorticity (the angular momentum of vortex) parallel to the chirality, and the other type has vorticity anti-parallel to the chirality. In the present paper, we call the former ``parallel vortex'' and the latter ``anti-parallel vortex''. In the Ginzburg-Landau(GL) theory, an anti-parallel vortex was shown to be more stable than a parallel vortex\cite{Heeb1999}.

To solve \eqref{Eq:Eilenberger}, we used so-called Riccati-parametrization method\cite{Nagato1993,Schpohl1995} and solved the parametrized differential equation with a 4th- and 5th-order adaptive Runge-Kutta method. We used the cylindrical coordinate system and took 48 equally spaced points on the azimuthal coordinates. To improve the accuracy of numerical integration of the free-energy, we used the composite Gauss-Lobatto quadrature to choose discrete points \(r_i\) on the radial line. We divided the closed interval \([0,2.4]\) into 16 subintervals, applied the 7-points Gauss-Lobatto formula to each subinterval and obtained 97 discrete points \(x_i\), and changed the variable as \(r_i = x_i+x_i^2/2+x_i^3/3+x_i^4/4\) in order to make the sampling points denser near the center and more sparse far from the vortex. We calculated the self-energy from the quasiclassical Green's functions via \eqref{Eq:Eliashberg} and iterated the above until the self-energy sufficiently converged.
After obtaining converged solution, we calculated the free-energy of vortices using \eqref{Eq:free-energy-super-normal}. We changed initial profiles of the vortices so that the initial dominant- and induced- vortices were at separate positions, and repeated the same procedure as the above. Finally, we compared the resultant profiles and their free-energies to discuss stability.

\section{Results and Discussion}

\begin{figure}
	\centering
	\includegraphics[width=.4\textwidth]{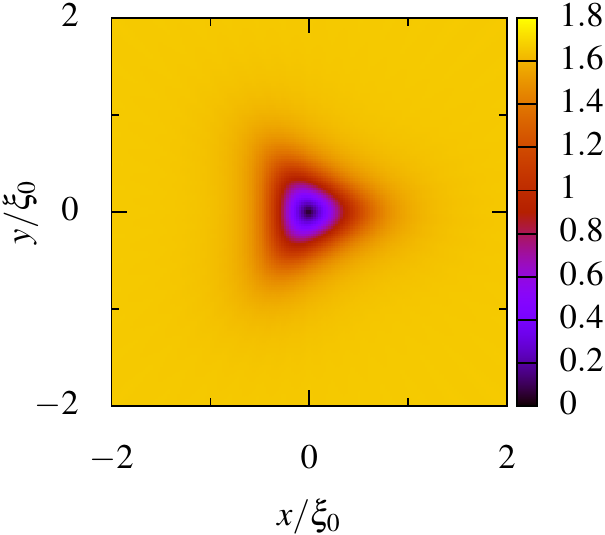}
	\hspace*{.05\textwidth}
	\includegraphics[width=.4\textwidth]{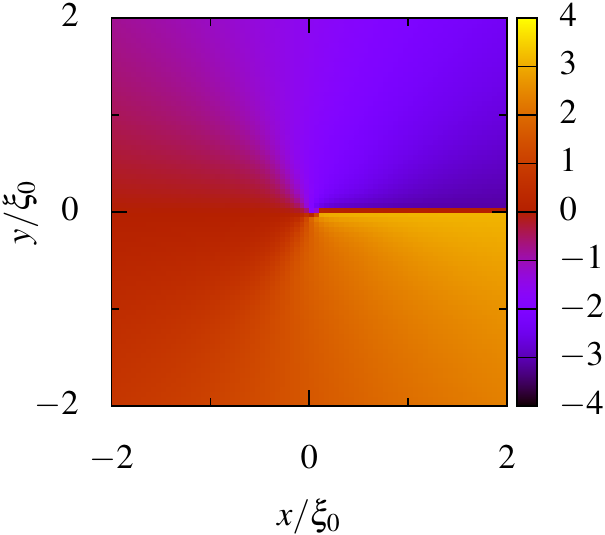}

	\includegraphics[width=.4\textwidth]{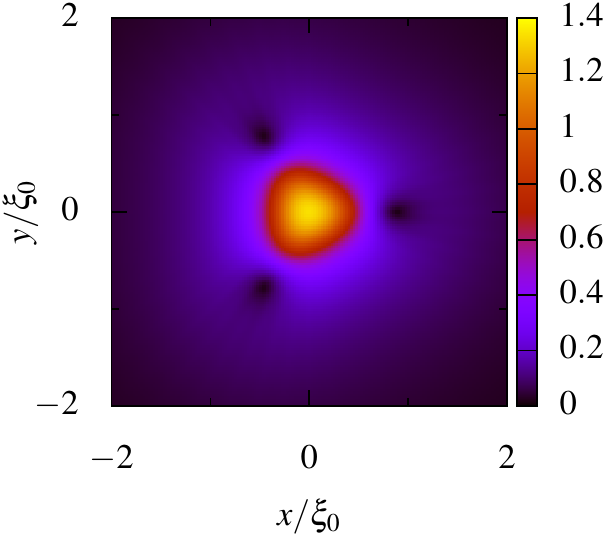}
	\hspace*{.05\textwidth}
	\includegraphics[width=.4\textwidth]{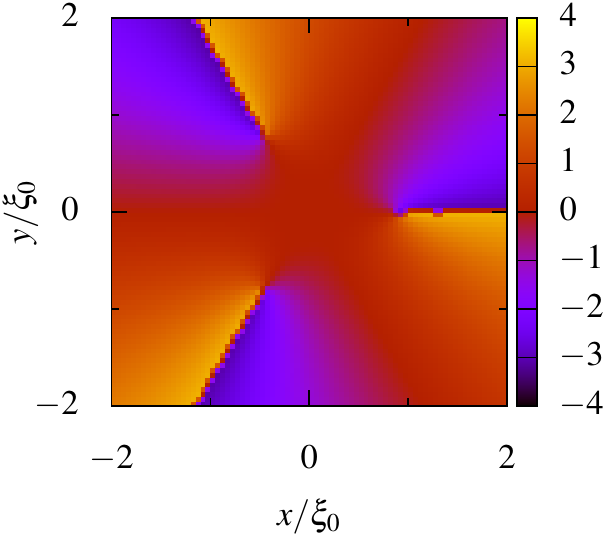}
	\caption{Profile of the off-diagonal part of the self-energy of the parallel vortex at \(T=0.3T_{\text{c}}\) and \(\hbar\epsilon_n=\cpi \ck T\). Left-top: amplitude of dominant part (\(h(\ci\epsilon_n)(\check\Sigma_{+})_{12}/T_{\text{c}}\)), right-top: phase of dominant part (\(\arg(\check\Sigma_{+})_{12}\)), left-down: amplitude of induced part (\(h(\ci\epsilon_n)(\check\Sigma_{-})_{12}/T_{\text{c}}\)), right-down: phase of induced part (\(\arg(\check\Sigma_{-})_{12}\)). (Color figure online)}
	\label{Fig:profile-of-non-axisymmetric-parallel-vortex}
\end{figure}

\begin{figure}
	\centering
	\includegraphics[width=.4\textwidth]{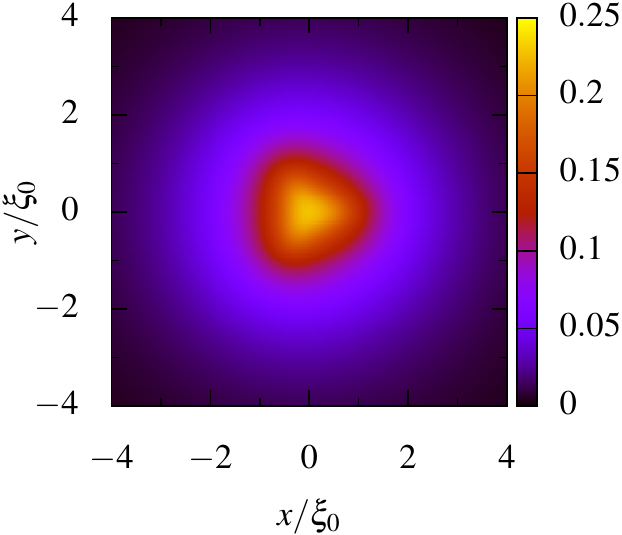}
	\hspace*{.05\textwidth}
	\includegraphics[width=.35\textwidth]{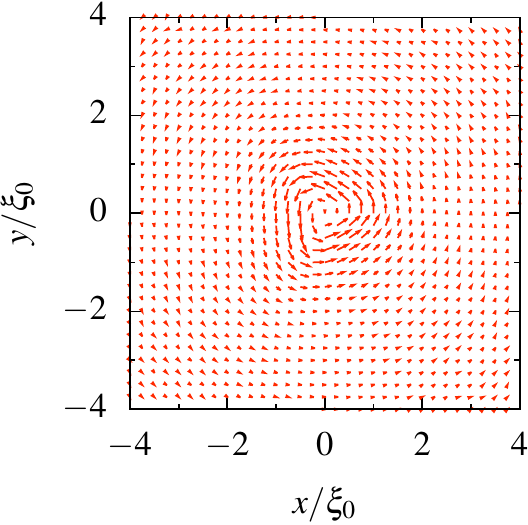}
	\caption{Electromagnetic quantities around the non-axisymmetric parallel vortex. Left: magnetic field, right: electric current density. (Color figure online)}
	\label{Fig:current-in-non-axisymmetric-parallel-vortex} 
\end{figure}

\begin{figure}
	\centering
	\includegraphics[width=.6\textwidth]{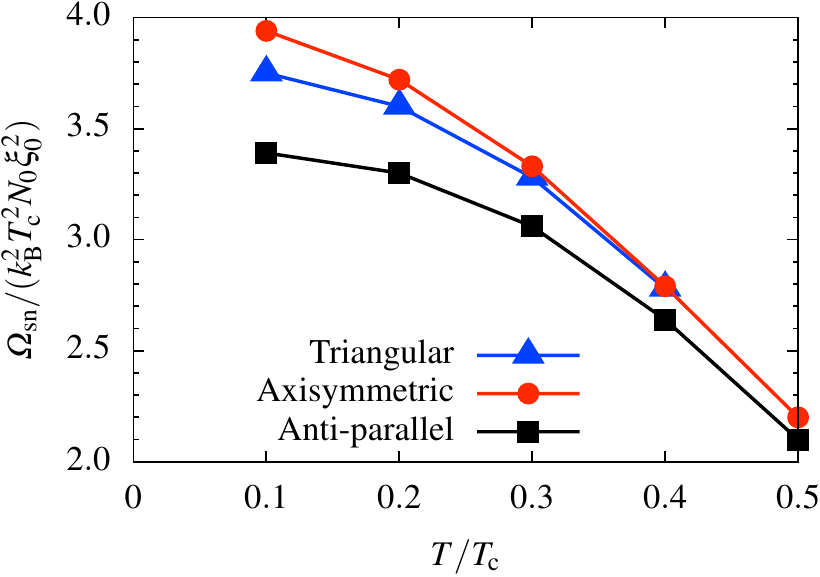}
	\caption{Free energy of each vortex. Red-circle: circular parallel vortex, blue-triangle: triangular parallel vortex, black-square: circular anti-parallel vortex. (Color figure online)}
	\label{Fig:free-energy-for-parallel-vortex}
\end{figure}

As for the anti-parallel vortices, we only obtained circular axisymmetric vortices for all temperature that we studied (\(T/T_{\text{c}}=0.1\), \(0.2\), \(0.3\), \(0.4\), and \(0.5\)), regardless of the initial profiles; in this case, the strong-coupling effect just modifies the shape of the vortex.

On the other hand, at moderately low temperatures, a non-axisymmetric solution emerges for parallel vortices, when the initial vortex sufficiently breaks the axisymmetry. 
%
%This dependence on initial conditions indicates that one of the solutions are a metastable state, and both vortices can exist in the same sample.
%
Figure \ref{Fig:profile-of-non-axisymmetric-parallel-vortex} shows the non-axisymmetric profile of dominant and induced parts of off-diagonal self-energy at \(T/T_{\text{c}}=0.3\) and \(\hbar\epsilon_n = \cpi \ck T\). There the vortex of dominant component forms triangle and those of the induced component split into three. Figure \ref{Fig:current-in-non-axisymmetric-parallel-vortex} shows the current density around the vortex. We can confirm that the electromagnetic quantities also break the axisymmetry.

When we calculated parallel vortices at \(T/T_{\text{c}} = 0.5\), we found only an axisymmetric vortex:  both circular and non-circular initial configurations of self-energy produce the same result. We thus conclude that unusual parallel vortices may not exist at high temperatures.
%This is consistent with the GL calculation, where we can obtain only an axisymmetric result.

%At \(T/T_{\text{c}}=0.1\), contrary to the situation in high-temperature, the self-consistent self-energy fallen into non-axisymmetric one: axisymmetric vortex becomes unstable for low-temperature.
%
In Figure \ref{Fig:free-energy-for-parallel-vortex}, we plot the free-energy of each vortex. We can see that at low temperatures, the non-axisymmetric vortex is more stable than symmetric one.
We note that the symmetric anti-parallel vortex is more stable than the non-axisymmetric parallel vortex, at least under the parameters in this study.

There are many studies of non-axisymmetric vortices in spin triplet superfluids or superconductors with an isotropic Fermi surface. However, many of them have targeted vortices in the superfluid ${}^3$He-B\cite{Thuneberg1986,Thuneberg1987,Salomaa1986,Fogelstroem1995,Tsutsumi2015,Silaev2015,Kondo1991}, or an \textit{f}-wave superconductor similar to the ${}^3$He-B\cite{Tsutsumi2012}; these studies therefore cannot be compared with our work directly.

Tokuyasu, \textit{et al.} have studied two-dimensional chiral \textit{p}-wave superconductor within the GL theory in the weak- to strong-coupling regimes\cite{Tokuyasu1990}. They have reported that non-axisymmetric vortices can emerge in some non-weak-coupling coefficients. However, the coefficients of the GL-functional of our target system fall into the same ones that we obtain in the weak-coupling limit (the \(\beta\) parameter in Ref.~\cite{Tokuyasu1990} is 0.5 in our system). Thus, the origin of non-axisymmetric vortices in our work is different from that of the previous work.
This is also consistent with the fact that non-axisymmetric vortices only exist at low temperatures in the present work.
Aoyama and Ikeda have reported that a vortex of ${}^3$He-A can be non-axisymmetric under the existence of anisotropic scatterers\cite{Aoyama2010-JLTP,Aoyama2010-PRB}. Their model is different from ours, and the relationship between their and our results considered an important but remaining issue.

\section{Conclusion}
In this study, we numerically found that a non-axisymmetric vortex metastably exists in strong-coupling chiral \textit{p}-wave superconductors. This anomalous vortex is more stable than the axisymmetric parallel one at sufficiently low temperatures, but symmetric anti-parallel vortex is still most stable.
The emergence of this anomalous vortex is a consequence of the strong-coupling effect because we did not obtain such a vortex with the conventional weak-coupling gap equation.
To clarify the underlying energetics that makes the non-axisymmetric vortex metastable is an interesting issue.
%We have not determined the physics behind this phenomena yet, and it should be desired to study other quantities to identify the physics of this system.
The total phase diagram of this system is also left as a future issue.

\begin{acknowledgements}
We thank J.\ A.\ Sauls and Y.\ Tsutsumi for helpful discussions.
This work was supported by JSPS KAKENHI Grant Number 15K05160.
\end{acknowledgements}


\begin{thebibliography}{00}
\bibitem{Migdal1958} A.\ B.\ Migdal, Sov.\ Phys.\ JETP \textbf{7}, 996 (1958).
\bibitem{Eliashberg1960} G.\ M.\ \'{E}liashberg, Sov.\ Phys.\ JETP \textbf{11}, 696 (1960).
\bibitem{Morel1962} P.\ Morel, P.\ W.\ Anderson, Phys.\ Rev.\ \textbf{125}, 1263 (1962).

\bibitem{Scalapino1969} D.\ J.\ Scalapino, in \textit{Superconductivity}, edited by R.\ D.\ Parks (Marcel Dekker, Inc., New~York, 1969) p.~449.
\bibitem{McMillan1969} W.\ L.\ McMillan, J.\ M.\ Rowell, in \textit{Superconductivity}, edited by R.\ D.\  Parks (Marcel Dekker, Inc., New~York, 1969) p.~561.
\bibitem{Carbotte1990} J.\ P.\ Carbotte, Rev.\ Mod.\ Phys.\ \textbf{62}, 1027 (1990).


\bibitem{Marsiglio1991} F.\ Marsiglio, J.\ P.\ Carbotte, Phys.\ Rev.\ B \textbf{43}, 5355 (1991).
\bibitem{Combescot1995} R.\ Combescot, Phys.\ Rev.\ B \textbf{51}, 11625 (1995).


\bibitem{Maeno1994} Y.\ Maeno, H.\ Hashimoto, K.\ Yoshida, S.\ Nishizaki, T.\ Fujita, J.\ G.\ Bednorz, F.\ Lichtenberg, Nature (London) \textbf{372}, 532 (1994).
\bibitem{Mackenzie2003} A.\ P.\ Mackenzie, Y.\ Maeno, Rev.\ Mod.\ Phys.\ \textbf{75}, 657 (2003).
\bibitem{Sigrist2005} M. Sigrist, Prog.\ Theor.\ Phys.\ Suppl.\ \textbf{160}, 1 (2005).
\bibitem{Maeno2012} Y.\ Maeno, S.\ Kittaka, T.\ Nomura, S.\ Yonezawa, K.\ Ishida, J.\ Phys.\ Soc.\ Jpn.\ \textbf{81}, 011009 (2012).

\bibitem{Volovik1999} G.\ E.\ Volovik, JETP Lett.\ \textbf{70}, 609 (1999).
\bibitem{Matsumoto2000} M.\ Matsumoto, M.\ Sigrist, Physica B \textbf{281\&282}, 973 (2000).
\bibitem{Kato2000} Y.\ Kato, J.\ Phys.\ Soc.\ Jpn.\ \textbf{69}, 3378 (2000).
\bibitem{Hayashi2005} N.\ Hayashi, Y.\ Kato, M.\ Sigrist, J.\ Low Temp.\ Phys.\ \textbf{139}, 79 (2005).
\bibitem{Tanuma2009} Y.\ Tanuma, N.\ Hayashi, Y.\ Tanaka, A.\ A.\ Golubov, Phys.\ Rev.\ Lett.\ \textbf{102}, 117003 (2009).
\bibitem{Eschrig2009} M.\ Eschrig, J.\ A.\ Sauls, New.\ J.\ Phys.\ \textbf{11}, 075008 (2009).
%\bibitem{Ichioka2012} M.\ Ichioka, K.\ Machida, J.\ A.\ Sauls, J.\ Phys.\ Conf.\ Ser.\ \textbf{400}, 022031 (2012).
\bibitem{Kurosawa2013} N.\ Kurosawa, N.\ Hayashi, E.\ Arahata, Y.\ Kato, J.\ Low Temp.\ Phys.\ \textbf{175}, 365 (2013).
\bibitem{Kurosawa2015} N.\ Kurosawa, N.\ Hayashi, Y.\ Kato, J.\ Phys.\ Soc.\ Jpn.\ \textbf{84}, 114710 (2015).
\bibitem{Tanaka2016} K.\ K.\ Tanaka, M.\ Ichioka, S.\ Onari, Phys.\ Rev.\ B \textbf{93}, 094507 (2016).

\bibitem{Tanaka2012} Y.\ Tanaka, M.\ Sato, N.\ Nagaosa, J.\ Phys.\ Soc.\ Jpn.\ \textbf{81}, 011013 (2012).


\bibitem{Eilenberger1968} G.\ E.\ Eilenberger, Z.\ Phys.\ \textbf{214}, 195 (1968).
\bibitem{Larkin1968} A.\ I.\ Larkin, Yu.\ N.\ Ovchinnikov, Sov.\ Phys.\ JETP \textbf{28}, 1200 (1968).

\bibitem{Park2008} W.\ K.\ Park, J.\ L.\ Sarrao, J.\ D.\ Thompson, L.\ H.\ Greene, Phys.\ Rev.\ Lett.\ \textbf{100}, 177001 (2008).

\bibitem{Serene1983} J.\ W.\ Serene, D.\ Rainer, Phys.\ Rep.\ \textbf{101}, 221 (1983).
\bibitem{Thuneberg1984} E.\ V.\ Thuneberg, J.\ Kurkij\"{a}rvi, D.\ Rainer, Phys.\ Rev.\ B \textbf{29}, 3913 (1984).

\bibitem{Heeb1999} R.\ Heeb, D.\ F.\ Agterberg, Phys.\ Rev.\ B \textbf{59}, 7076 (1999). 

%\bibitem{Lee1988} W.\ Lee, D.\ Rainer, Z.\ Phys.\ B \textbf{73}, 149 (1988).
\bibitem{Nagato1993} Y.\ Nagato, K.\ Nagai, J.\ Hara: J.\ Low Temp.\ Phys.\ \textbf{93}, 33 (1993).
\bibitem{Schpohl1995} N.\ Schophol, K.\ Maki, Phys.\ Rev.\ B \textbf{52}, 490 (1995).


\bibitem{Thuneberg1986} E.\ V.\ Thuneberg, Phys.\ Rev.\ Lett.\ \textbf{56}, 359 (1986).
\bibitem{Thuneberg1987} E.\ V.\ Thuneberg, Phys.\ Rev.\ B \textbf{36}, 3583 (1987).
\bibitem{Salomaa1986} M.\ M.\ Salomaa, G.\ E.\ Volovik, Phys.\ Rev.\ Lett.\ \textbf{56}, 363 (1986).
\bibitem{Fogelstroem1995} M.\ Fogelstr\"om, J.\ Kurkij\"arvi, J.\ Low Temp.\ Phys.\ \textbf{98}, 195 (1995).
\bibitem{Tsutsumi2015} Y.\ Tsutsumi, T.\ Kawakami, K.\ Shiozaki, M.\ Sato, K.\ Machida, Phys.\ Rev.\ B \textbf{91}, 144504 (2015).
\bibitem{Silaev2015} M.\ A.\ Silaev, E.\ Thuneberg, M.\ Fogelstr\"om, Phys.\ Rev.\ Lett.\ \textbf{115}, 235301 (2015).

\bibitem{Kondo1991} Y.\ Kondo, J.\ S.\ Korhonen, M.\ Krusius, V.\ V.\ Dmitriev, Y.\ M.\ Mukharsky, E.\ B.\ Sonin, G.\ E.\ Volovik, Phys.\ Rev.\ Lett. \textbf{67}, 81 (1991).



\bibitem{Tsutsumi2012} Y.\ Tsutsumi, K.\ Machida, T.\ Ohmi, M.\ Ozawa, J.\ Phys.\ Soc.\ Jpn.\ \textbf{81}, 074717 (2012).
%\bibitem{Mermin1976} N.\ D.\ Mermin, Tin-Lun Ho, Phys.\ Rev.\ Lett.\ \textbf{36}, 594 (1976).


\bibitem{Aoyama2010-JLTP} K.\ Aoyama, R.\ Ikeda, J.\ Low Temp.\ Phys.\ \textbf{158}, 404 (2010).
\bibitem{Aoyama2010-PRB} K.\ Aoyama, R.\ Ikeda, Phys.\ Rev.\ B \textbf{82}, 144514 (2010).


\bibitem{Tokuyasu1990} T.\ A.\ Tokuyasu, D.\ W.\ Hess, J.\ A.\ Sauls, Phys. Rev. B \textbf{41}, 8891 (1990).

\end{thebibliography}
\end{document}